\documentclass[%aip,
aps,
 amsmath,amssymb,
 reprint,%
]{revtex4-2}
\usepackage{float}
\usepackage{graphicx}% Include figure files
\usepackage{dcolumn}% Align table columns on decimal point
\usepackage{bm}% bold math
\usepackage[T1]{fontenc}
\usepackage[english]{babel}
\usepackage{mathptmx}
\usepackage{xcolor}
\usepackage{amssymb}

\usepackage{pgfplots}
\pgfplotsset{compat=1.15}

\newcommand{\be}{\begin{equation}}
\newcommand{\ee}{\end{equation}}
\newcommand{\bea}{\begin{eqnarray}}
\newcommand{\eea}{\end{eqnarray}}

\begin{document}

 % \preprint{AIP/123-QED}

\title[]{Theoretical Models for Tension-Dependent DNA Looping Time}
% Force line breaks with \\

\author{Wout Laeremans}
\affiliation{Soft Matter and Biological Physics, Department of Applied Physics and Science Education, and Institute for Complex Molecular Systems,
Eindhoven University of Technology, P.O. Box 513, 5600 MB Eindhoven, Netherlands}
\author{Wouter G. Ellenbroek}
\email{w.g.ellenbroek@tue.nl}
\affiliation{Soft Matter and Biological Physics, Department of Applied Physics and Science Education, and Institute for Complex Molecular Systems,
Eindhoven University of Technology, P.O. Box 513, 5600 MB Eindhoven, Netherlands}

\date{\today}

\begin{abstract}
The influence of tension on DNA looping has been studied both experimentally and theoretically in the past. However, different theoretical models have yielded different predictions, leaving uncertainty about their validity. We briefly review the predictions of those models and propose a novel model that demonstrates exceptional agreement with simulations for long semiflexible chains. Additionally, we elucidate the relationship between our result and that of the previously proposed two-state model, highlighting the distinct interpretative approach that underpins our framework. Our findings offer predictive insights that pave the way for future experimental validation.
\end{abstract}

\maketitle

%%%%%%%%%%%%%%%%%%%%%%%%%%%%%%%%%%%%%%%%%%%%%%%%%%%%%%%%%%%%%%%%%%%%%%%%
% Introduction                                                         %
%%%%%%%%%%%%%%%%%%%%%%%%%%%%%%%%%%%%%%%%%%%%%%%%%%%%%%%%%%%%%%%%%%%%%%%%
\section{Introduction}
\begin{figure}[t]
    \centering
    \includegraphics[width=\linewidth]{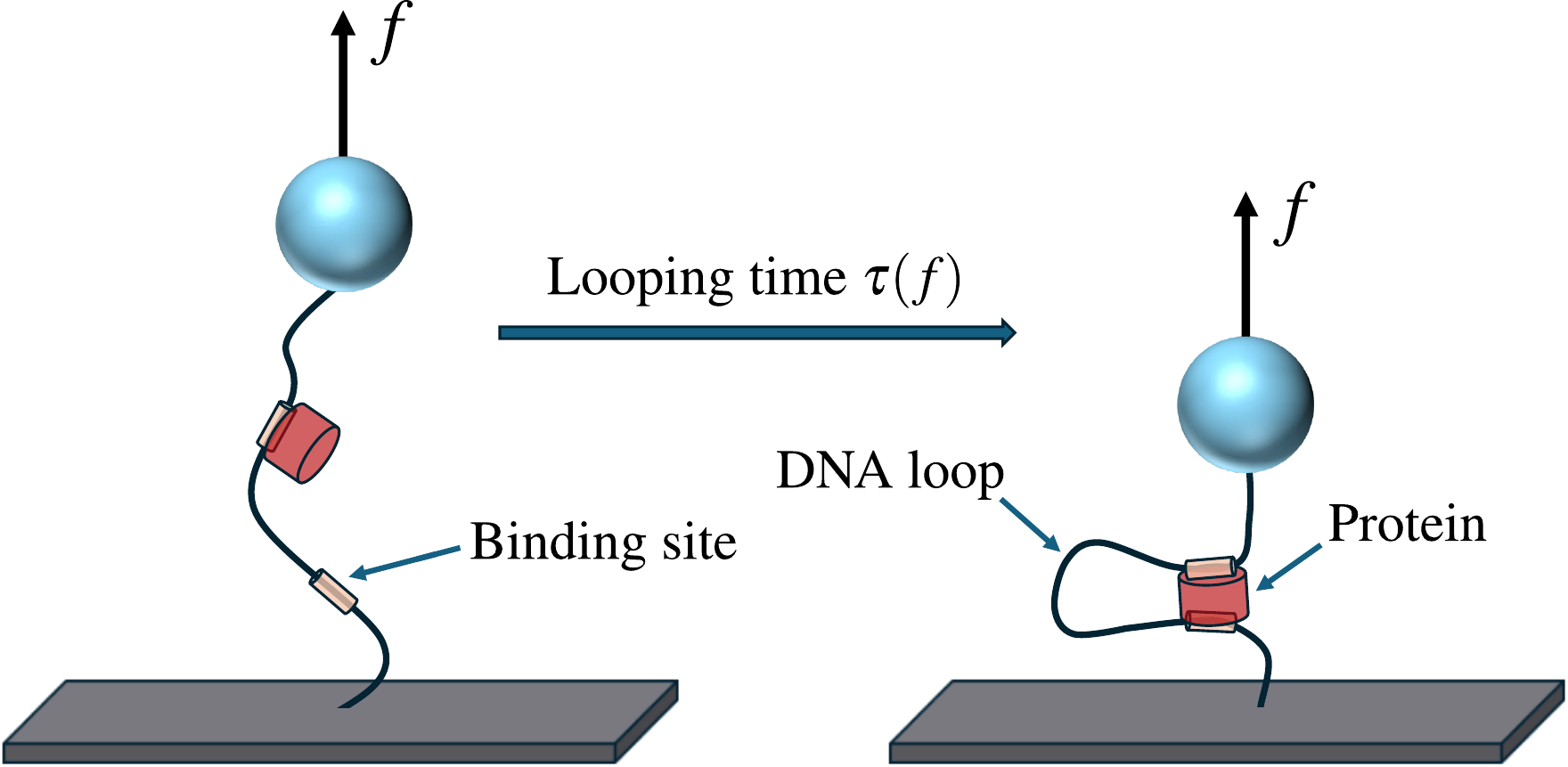}
\caption{Protein-mediated DNA looping occurs when two distal binding sites come into close contact and become bound by a protein or protein-complex. Using a tethered bead, tension can be applied resulting in a force-dependent looping time $\tau(f)$. Such experiments have been performed in Ref.~\onlinecite{chen2010protein} and show that forces below a piconewton can alter the looping time by an order of magnitude.}
\label{fig:dnamediatlooping} 
\end{figure}

DNA looping is a topic of high interest in the field of biology, as it is important for gene regulation~\cite{schleif1992dna, allemand2006loops, milstein2011role}, as well as DNA recombination, packaging, and many more~\cite{matthews1992dna}. Protein-mediated DNA loops are mainly driven by thermal fluctuations with forces around $\sim 80$ fN~\cite{chen2010femtonewton}. However, within the highly dynamic and out-of-equilibrium environment of a biological cell, DNA is continually subjected to piconewton-scale forces from its intracellular surroundings, which can exceed those that stem from typical thermal fluctuations by an order of magnitude~\cite{chen2010femtonewton, rivas2004life, gallet2009power}. This effect of tension on the looping time was measured using optical trapping, and it became clear that forces less than a piconewton can increase the looping time by an order of magnitude~\cite{chen2010protein} (see Fig.~\ref{fig:dnamediatlooping} for an illustration). 

This dependence of the looping time on the tension in the DNA has been investigated theoretically in the past by analyzing the cyclization process of a semiflexible polymer. Blumberg et al.~\cite{blumberg2005disruption} started this exploration, studying protein-mediated DNA looping as a two-state system under the assumption of detailed balance. Later, Shin et al.~\cite{shin2012effects} investigated the same topic as a barrier escape problem. In the low-force regime ($f < 80$ fN), which is particularly relevant in biological contexts, the theories disagree in their predictions for the force-dependence of the loop formation time. The first theory suggests that the looping time increases exponentially in $f^2$, while the second theory indicates an almost exponential increase in $f$. Up to this day, it remains unclear which --- if either --- is correct. Moreover, in this low force regime there is no experimental data available to put these theories to the test. At higher forces, one can show that the barrier escape approach of Shin et al.~\cite{shin2012effects} is in very good agreement with the available experimental data of Chen et al.~\cite{chen2010protein}, while the two-state model of Blumberg et al.~\cite{shin2012effects} is not, as we discuss in Section~\ref{sec:review}.

In this Letter, we demonstrate that this good agreement between experiment and the barrier escape theory should be expected to break down for smaller forces, and furthermore that the assumptions within this theory become inaccurate when considering longer DNA strands.
We then present a novel third approach, which agrees extremely well with simulations for long chains under any force, a regime that could not be explained by prior theories. We explain in what sense our approach is related to the two-state model, and recover, in our model, the quadratic-exponential growth of the two-state model at the lowest forces. This offers new insights into looping of semiflexible polymers and provides a straightforward strategy for calculating looping times. Furthermore, we provide experimentally verifiable predictions and thus pave the way for future experimental investigations.

\section{Review of prior theories}
\label{sec:review}
Theoretical models for protein-mediated DNA looping approximate the system simply as the cyclization of a semiflexible chain, with the ends representing the binding sites. Shin et al.~\cite{shin2012effects} studied the looping process as a barrier escape problem: Imagining one of the ends of the polymer as fixed, the other can be seen as diffusing in an effective free energy landscape $F(r)$, with $r$ the end-to-end distance. The looping time then becomes the mean first passage time for this diffusing end to reach a certain capture radius $r_\mathrm{c}$ (distance between the binding sites when bound by protein) and reads~\cite{shin2012effects}
\begin{align}
    \tau_\text{Shin}(f) &\sim \frac{1}{Z} \int_{r_\mathrm{c}}^{L}dr \int_{r_\mathrm{c}}^{r}  dr'\int_{r'}^{L} dr'' e^{-\beta \left[ F(r) - F(r') + F(r'') \right]}, \label{eq:tripple_integral}
\end{align}
with $Z = \int_{r_\mathrm{c}}^{L} e^{-\beta F(r)} \ dr$ the partition function, $\beta$ the inverse temperature and $L$ the total DNA length between the two binding sites. Although in an optical tweezer experiment, the force is applied in a specific direction, one can assume the end-to-end distance to align with the force at sufficiently high tension, resulting in a free energy of the form $\beta F_\text{Shin}(r) = -\ln\left[ P(r, f=0) \right] - \beta f r$. For $P(r, f=0)$, Shin et al.~\cite{shin2012effects} used the mean field result for a semiflexible chain~\cite{grosberg1998theoretical, shin2012effects}
\begin{align}
    P_\text{MF}(r, f=0) &\sim r^2 \left[ 1 - \left( \frac{r}{L} \right)^2 \right]^{-9/2} e^{\frac{-3L}{4l_P \left[ 1 - \left( \frac{r}{L} \right)^2 \right]}}. \label{eq:MF_probShin}
\end{align}
By contrast, Blumberg et al.~\cite{blumberg2005disruption}  approached the looping process as a two-state system: Either the polymer is looped or it is not. Denoting the energy difference between the looped and unlooped state as $\Delta G(f) = F\left( x_\mathrm{c} \right) - F\left( \langle x(f) \rangle \right)$ with $x$ the projection of the end-to-end distance along the force, $\langle x \rangle$ its average value and $x_\mathrm{c}$ the capture distance such that a loop obeys $x \leq x_\mathrm{c}$, they argued that~\cite{blumberg2005disruption}
\begin{align}
    \tau_\text{Blum.}(f) &\sim e^{\beta \Delta G(f)}. \label{eq:Blumbergeq}
\end{align}
By integration of the Marko-Siggia force-extension relation of a semiflexible chain~\cite{marko1995stretching}, they found the following free energy~\cite{blumberg2005disruption, laeremans2024looping}
\begin{align}
    \beta F_\text{Blum.}(x) &= \frac{1}{l_P} \left[ -\frac{L^2}{4(x - L)} + \frac{x^2}{2L} - \frac{x}{4} \right] - \beta f x, \label{eq:blumfree}
\end{align}
where the force is chosen to be in the $x$-direction. Although both approaches are different, one could in fact see the two-state model as an approximation of the barrier escape approach. In a two-state model~\cite{blumberg2005disruption}, one neglects the transition path, which is integrated over in the approach of Shin et al.~\cite{shin2012effects}. Mathematically, this makes Eq.~\ref{eq:Blumbergeq} a first order saddle-point approximation of Eq.~\ref{eq:tripple_integral} using the same free energy. 

\begin{figure}[t]
    \centering
    \includegraphics[width=0.8\linewidth]{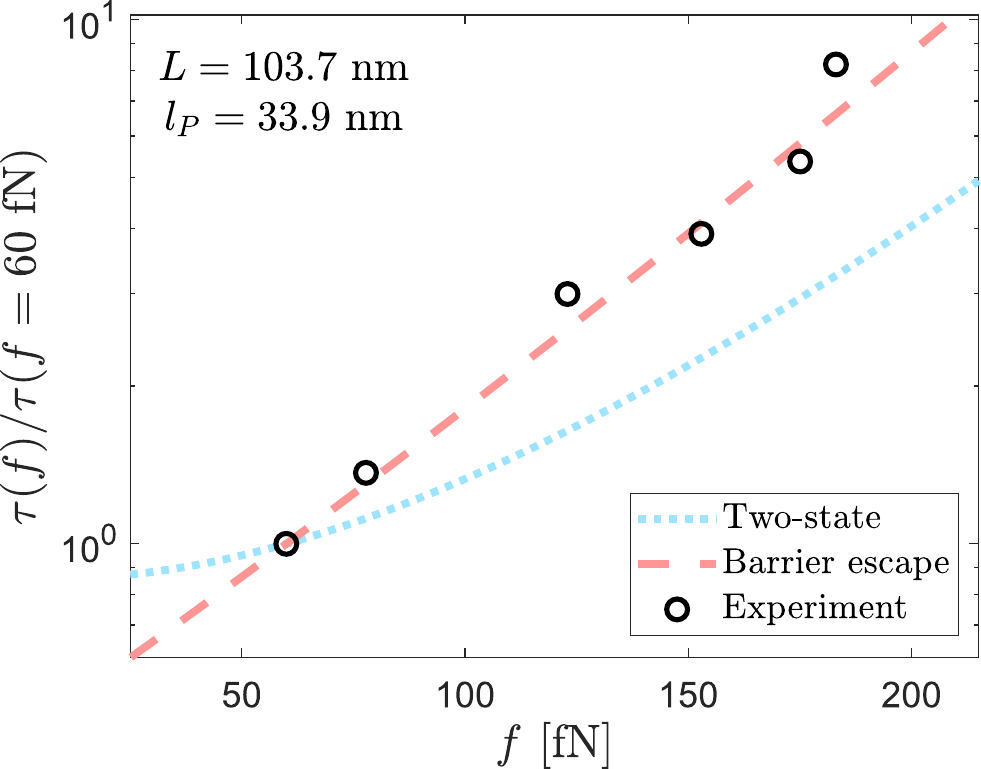}
\caption{Experimental data of Chen et al.~\cite{chen2010protein} measured using an optical tweezer setup, compared to the theories of Blumberg et al.~\cite{blumberg2005disruption} (two-state) and of Shin et al.~\cite{shin2012effects} (barrier escape). We used $L = 103.7$ nm, $r_\mathrm{c} = x_\mathrm{c} = 5$ nm, $\beta = 4114^{-1}$ fN $\times$ nm and $l_P = 33.9$ nm. (The data was extracted from Ref.~\cite{chen2010protein} using PlotDigitizer. See Supplementary Material for additional information regarding the choice of parameters.)}
\label{fig:comp_exp} 
\end{figure}
In Fig.~\ref{fig:comp_exp}, we show as black circles the experimental data of Chen et al.~\cite{chen2010protein}, together with the prediction of Blumberg et al.~\cite{blumberg2005disruption} (two-state) and of Shin et al.~\cite{shin2012effects} (barrier escape). As is clear, the latter is in excellent agreement with the experimental data, while the former is not. We note however, that it is difficult to judge a theoretical model based on one experiment only. Therefore, we will compare to simulation data for varying polymer lengths and over a more extended force regime, once we have introduced our novel model.

\section{Theoretical model}
In an ergodic Markov chain with a stationary distribution, the mean recurrence time of a state is inversely proportional to the long-term probability of occupying that state~\cite{wolfowitz1951william}. This relationship arises naturally, as the stationary probabilities represent the average fraction of time the system spends in each state. Without tension, the looping time was also shown by Jun and coworkers to be inversely proportional to the equilibrium probability to form a loop~\cite{jun2003diffusion}. We confirm these findings in the Supplementary Material. Recent work reached the same conclusion for a freely jointed chain under tension~\cite{laeremans2024looping}. The latter also showed that the approach of Shin et al.~\cite{shin2012effects} (Eq.~\ref{eq:tripple_integral}) might fail due to local equilibrium not being satisfied, while the inverse scaling with the looping probability still holds~\cite{laeremans2024looping}. Here, we extend these recent findings to semiflexible chains, connecting them to the work of Jun. We start by demonstrating that the inverse relationship between looping time and looping probability reproduces the same scaling in $f$ as the two-state model of Blumberg et al.~\cite{blumberg2005disruption} for $x_\mathrm{c} \rightarrow 0$. In the process, we recover the prediction that the looping time grows exponentially in $f^2$ for small values of $f$.

\begin{figure*}[t]
\centering
\includegraphics[width=\textwidth]{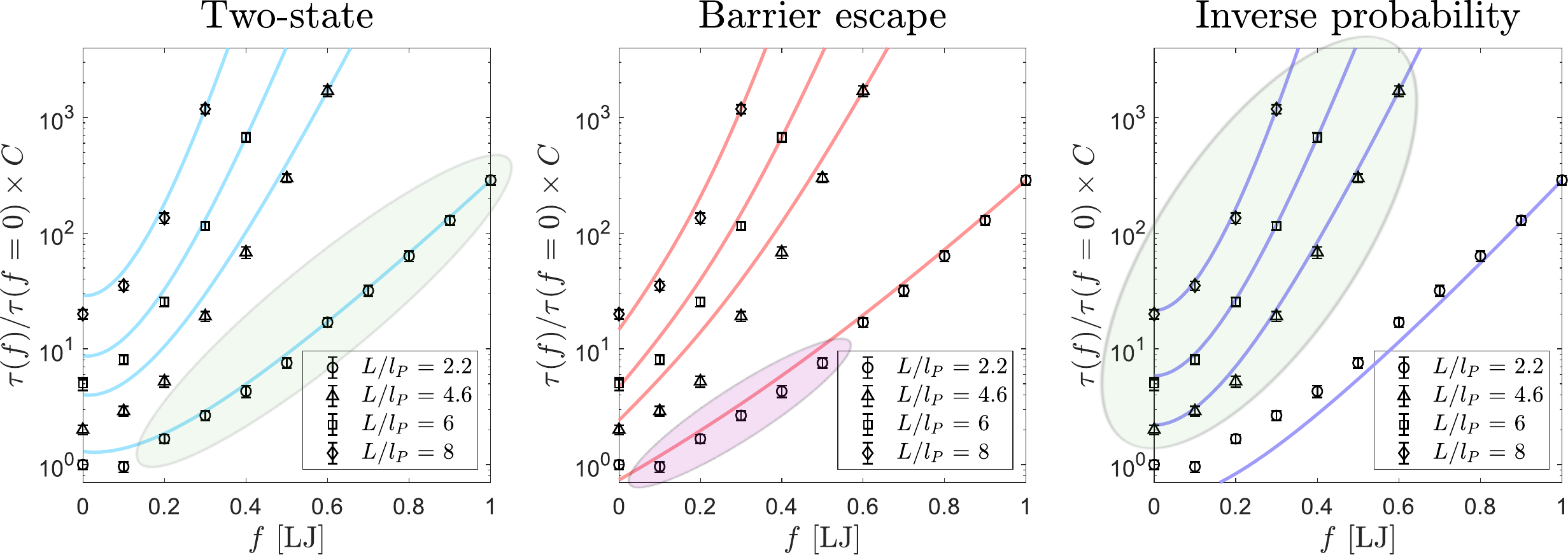}
  \caption{Three different theories (from left to right): Blumberg et al.~\cite{blumberg2005disruption} (two-state model), Shin et al.~\cite{shin2012effects} (barrier escape approach) and our model, using the inverse scaling with the looping probability. Those are compared with simulations for a semiflexible chain, using the parameters (in LJ dimensionless units): $\beta = b = r_\mathrm{c} = \gamma = 1$, $K = 100$, $\kappa = 5$ and $C = 1, 2, 5, 20$ for $N = 11, 23, 30, 40$ respectively. As the theories only provide a scaling in $f$, we made them coincide with the simulation for the most upper data point at every length. One can observe that for a short chain ($L/l_P = 2.2$) at high forces ($f > 0.2$), the two-state model makes a good prediction. For long chains, both prior theories fail, while our model on the right makes consistently a very good prediction on the whole force regime. The green areas highlight where good correspondence is found between theory and simulation. The pink area roughly corresponds to the region probed in the experiment of Chen et al.~\cite{chen2010protein} shown in Fig.~\ref{fig:comp_exp}.}
  \label{fig:comp_sim} 
\end{figure*}

The derivation in this section does not make any assumptions on the specific polymer model and therefore holds for both flexible and semiflexible chains. Very generally, we denote the free energy for a polymer under tension $F$ in function of the end-to-end vector $\vec{r}$ and applied force $f$ as $F(\vec{r}, f)$ and note that the probability of observing a specific $\vec{r}$ in equilibrium can be written as~\cite{laeremans2024looping}
\begin{align}
    P(\vec{r},f) &= \frac{e^{-\beta F(\vec{r},f)}}{Z(f)} = \frac{e^{-\beta F(\vec{r}, f=0) + \beta f x}}{Z(f)},
\end{align}
where we again take the force as applied in the $x$-direction. As we are interested in the scaling in $f$, we can write down
\begin{align}
    P(\vec{r},f) &\sim \frac{e^{\beta f x}}{Z(f)},
\end{align}
as $F(\vec{r}, f=0)$ does not depend on $f$. When a loop is formed, $x$ is close to zero, by which the exponential factor can be neglected, resulting in
\begin{align}
    P_\text{loop}(f) \sim \frac{1}{Z(f)}.
\end{align}
Moreover, we can use the thermodynamic relation $G(f) = -k_BT\ln[Z(f)]$ and note that $G(f) = F(\langle x(f) \rangle)$. This last statement explains that the free energy $F(x)$ in function of the extension $x$, yields the free energy $G(f)$ in function of the force $f$, when evaluated in the average extension $x = \langle x \rangle$. Hence, when the looping time is inversionally proportional to the looping probability, we have
\begin{align}
    \tau(f) \sim e^{-\beta F(\langle x(f) \rangle)}, \label{eq:Wout}
\end{align}
which can be evaluated fully analytically (see Supplementary Material). That is, we recover the result of Blumberg et al.~\cite{blumberg2005disruption} (Eq.~\ref{eq:Blumbergeq}) for $x_\mathrm{c} \rightarrow 0$. One should, however, note that we got this result from a different interpretation, namely as an inverse scaling with a looping probability, not as a two-state system. The validity of Eq.~\ref{eq:Wout} has been confirmed previously for a freely jointed chain in Ref.~\onlinecite{laeremans2024looping} and below, we will show it holds for long semiflexible chains as well.

\section{Simulation}
In order to validate the predictions of all three models, we perform simulations of a semiflexible chain undergoing Langevin dynamics. The polymer is represented by a bead-spring model with $N+1$ beads, a bond potential and an angle potential. The simulation uses dimensionless (Lennard-Jones) units, using an inverse temperature $\beta=1$ to set the unit of energy and a rest length for the springs of $b=1$ to set the unit of length. The friction coefficient is set to $\gamma = 1$, and the spring constant is taken to be $K = 100$, where we introduced a spring energy
\begin{align}
    U_\mathrm{S} &= \frac{K}{2}\sum_{i = 0}^{N-1} \left( |\vec{r}_{i+1} - \vec{r}_{i}| - b \right)^2,
\end{align}
with $\vec{r}_{i}$ the position vector of bead $i$. Furthermore, we included a bending potential with stiffness $\kappa = 5$ as
\begin{align}
    U_\mathrm{B} &= \frac{\kappa}{2} \sum_{i=2}^{N-1} \theta_i^2,
\end{align}
where $\theta_i$ represents the angle between bond vectors $i-1$ and $i$. The bending stiffness can be related to the persistence length~\cite{boal2012mechanics}, a commonly used quantifier for the bending stiffness of thermally fluctuating chains, as $l_P = \beta \kappa b$. As explained in Ref.~\onlinecite{shin2012effects}, these can be mapped to real units for DNA. Taking $b = 10$ nm and $l_p = 50$ nm, a force of $f = 1$ roughly corresponds to $400$ fN. For every simulation, we let the chain relax for a long time under a force $f$ applied to the end beads, after which we measure the looping time. A loop is defined as the mean first passage time for $r$ to become smaller than or equal to the capture radius $r_\mathrm{c}$ which we set to $r_\mathrm{c} = 1$, where the average is taken over 100 simulations for each force and length. Further details of the simulation procedure can be found in the Supplementary Material.

\section{Results}
In Fig.~\ref{fig:comp_sim}, we compare all three theories to simulation data, where the areas in green highlight good correspondence and the pink area marks the region that roughly corresponds to what is currently probed in optical tweezer experiments~\cite{chen2010protein}, as shown in Fig.~\ref{fig:comp_exp}. For short chains at high tension, the two-state model makes a very good prediction (green area left panel). This can be understood as follows: The two-state model is a first order saddle-point approximation of the barrier escape approach. Hence, it has to fulfill 1) local equilibrium, because all the degrees of freedom are mapped to a single reaction coordinate and 2) the barrier has to be high, as it is a saddle-point approximation. The former condition is satisfied for short chains as $l_P/L$ is large, hence fluctuations around the ground state are small and local equilibrium is reached faster. The latter condition is satisfied under high tension. 

The barrier escape approach makes a very similar prediction at the highest forces, but becomes less accurate more quickly when the force is lowered. The reason for this is that this method relies on constructing a free energy $F(r)$ in a single reaction coordinate $r$ (the end-to-end distance). At low forces, the end-to-end vector of the polymer is never perfectly aligned with the direction of the force, which implies one coordinate cannot capture the full behavior.
At high tension however, as explained in Sec.~\ref{sec:review}, the polymer will align itself, making the single reaction coordinate approximation perform better. The two-state model avoids the necessity of a 3-dimensional free energy landscape when this alignment is not present, by only depending on the energy difference, not on the transition path itself. Hence, its prediction remains valid for lower forces than in the case of the barrier escape approach. For long chains, both the two-state model and the barrier escape approach fail due to local equilibrium not being satisfied.

In Fig.~\ref{fig:comp_exp}, we did however observe that it was the barrier escape approach that made the best prediction. The experimental data was measured in the force range $60$ to $180$ fN, which roughly translates to $f$ going from $0.15$ to $0.45$ in our simulations. Furthermore, the length in the experiment is $L = 103.7$ nm, close to $L/l_P = 2.2$ in the simulations. In Fig.~\ref{fig:comp_sim}, this experimental regime more or less translates to the region marked in pink. Here, we can see a very similar scaling in $f$ between the barrier escape approach and simulation, but we note as well that the trend changes both for lower and higher forces. Hence, we believe the agreement is more by accident in this specific region.  

Finally, we show that our model using the inverse scaling with the looping probability consistently makes an excellent prediction for long chains, on the whole force regime (green area right panel). For the shortest chain ($N = 11$), we explicitly confirmed in the Supplementary Material that no finite size effects come into play due to bead-discritization. Therefore, the only reason for failure of our theory is that $r_\mathrm{c}$ is no longer small as compared to the total length. Reducing the capture radius however, would introduce other artifacts such as higher modes becoming more important~\cite{toan2008kinetics, laeremans2024looping}. For reasonable chain lengths ($L\gtrsim 4l_p$), our new model really complements to prior descriptions of protein-mediated DNA looping under tension. Note as well, that our model correctly predicts the quadratic-exponential increase in the looping time with respect to the force in the low force regime, similar to the two-state model of Blumberg et al.~\cite{blumberg2005disruption}. This is because, for low forces ($\beta f l_P \ll 1$) and long chains ($L/l_P \gg 1$), a semiflexible chain becomes an entropic spring~\cite{marantan2018mechanics}, hence the free energy is quadratic in $f$. This is not at all present in the barrier escape approach of Shin et al.~\cite{shin2012effects}, which predicts an almost exponential increase instead. We now resolve this debate, by showing that the quadratic-exponential scaling in $f$ is indeed correct for long semiflexible chains under low tension. Nevertheless, we stress again that our model is based on a different interpretation for why this is the case, than the two-state model of Blumberg et al.~\cite{blumberg2005disruption}.

\section{Conclusion}
In this work, we studied protein-mediated DNA looping under tension as a cyclization process of a semiflexible polymer. We first compared two existing theories, namely the barrier escape approach from Shin et al.~\cite{shin2012effects} and the two-state model from Blumberg et al.~\cite{blumberg2005disruption}, with the experimental data of Ref.~\onlinecite{chen2010protein}. From this, we found that only the former approach seems to agree. 

Next, we introduced a new theory, which starts from the idea that the looping time inversely scales with the equilibrium looping probability under an applied force $f$. From this we were able to show that this reduces to the two-state model of Blumberg et al.~\cite{blumberg2005disruption} in the limit of a vanishing capture radius. Therefore, our findings support the prediction of Ref.~\onlinecite{blumberg2005disruption} that in the smallest force regime, the looping time grows with the quadratic-exponential scaling in $f$. Nevertheless, we emphasize that the reduction of our model to a two-state framework arises purely as a mathematical coincidence. This highlights the potential risk of misinterpreting experimental data as fitting a two-state model when neglecting the size of the protein.

Comparing to simulations for a semiflexible chain, we found that for short chains under relatively high tension, the two-state model makes a good prediction, while for long chains, our inverse scaling with the looping probability performs well over the whole force range. This we explained by noticing that the two-state model is only valid under a local equilibrium assumption, which is valid for short chains and as it is a saddle-point approximation, it is only valid at high tension. Our model fails for short chains, as the capture radius should be small as compared to the total length. The barrier escape approach was found to not perform well in general. The reason for this, is that it relies on a free energy in a single reaction coordinate, which can only be approximately correct due to the asymmetry of the problem. Nevertheless, in the very specific force and length range Chen et al.~\cite{chen2010protein} used in their optical tweezer experiment, the barrier escape approach agrees the best. This we also confirmed with simulation (pink area in Fig.~\ref{fig:comp_sim}).

In summary we have shown that in order to describe the looping time of long semiflexible chains over a wide range of forces, our proposed method using the inverse of the looping probability performs best, while the more specific case of short chains under high tension is well described by the two-state model of Blumberg et al.~\cite{blumberg2005disruption}. Our simulations also resolve the discussion regarding the force dependence of the looping time in the lowest force regime, showing that it does increase exponentially in $f^2$ in case of long semiflexible chains.
%Although non of these conclusions have been validated experimentally, this work provides promising predictions and invites for additional research to confirm our findings.
We stress that our predictions should be within reach of experimental validation and are looking forward to the additional insights such experiments will yield.

\acknowledgments{We thank C. Storm for helpful suggestions. This research is financially supported by the Dutch Ministry of Education, Culture and Science (Gravity Program 024.005.020 – Interactive Polymer Materials IPM).}

\bibliography{references}

\appendix

\section*{SUPPLEMENTARY MATERIAL}
\subsection*{Parameters for the Optical Tweezer experiment}
\label{app:comp_exp}
In Ref.~\onlinecite{chen2010protein}, the looping time was measured experimentally using optical trapping. A 1316 bp dsDNA molecule
with two primary lac operators was used, of which one of the ends was attached to a polystyrene microsphere that could be trapped using a laser~\cite{chen2009stretching}. The lac operators were spaced 305 bp apart, which means that $L = 305 \ \text{bp} \times 0.34 \ \text{nm} = 103.7 \ \text{nm}$. For the temperature, we assume room temperature, giving $\beta = 4114^{-1} \ \text{fN} \times \text{nm}$. For the capture radius, we adopt the value of Shin et al.~\cite{shin2012effects}, namely $r_\mathrm{c} = 5$ nm. %We note however, that a slightly smaller or larger capture radius has a negligible effect. 

The final parameter to be discussed is the persistence length $l_P$. The usually accepted value~\cite{baum97} is $l_P = 45 \pm 5$ nm. In the work of Shin et al.~\cite{shin2012effects}, a persistence length $l_P = 50$ nm was used to compare to the experimental data of Ref.~\onlinecite{chen2010protein}. However, as the authors of the experiment explained in Ref.~\onlinecite{chen2009stretching}, when they fitted their experimentally measured force-extension relation to the semiflexible chain formalism with $l_P$ as a free parameter, they found $l_P = 33.9$ nm. This is much lower than the expected value. However, it has been observed in the past that when the persistence length is probed for a relatively short dsDNA molecule, it might be lower~\cite{seol2007elasticity}. From a theoretical point of view, one might also think of the persistence length as being length-scale dependent, a topic of ongoing research within the field of DNA mechanics~\cite{skoruppa2021length, segers2022mechanical, gutierrez2023coarse, laeremansDNA, skoruppa2024systematic}. As the theoretical predictions of this work fully rely on the semiflexible chain, we adopted the value of $l_P = 33.9$ nm in Fig.~\ref{fig:comp_exp}, as this best describes the experimental data~\cite{chen2009stretching}. 

\subsection*{Looping time without tension}
\label{app:loop_f0}
\begin{figure}[t]
    \centering
    \includegraphics[width=0.8\linewidth]{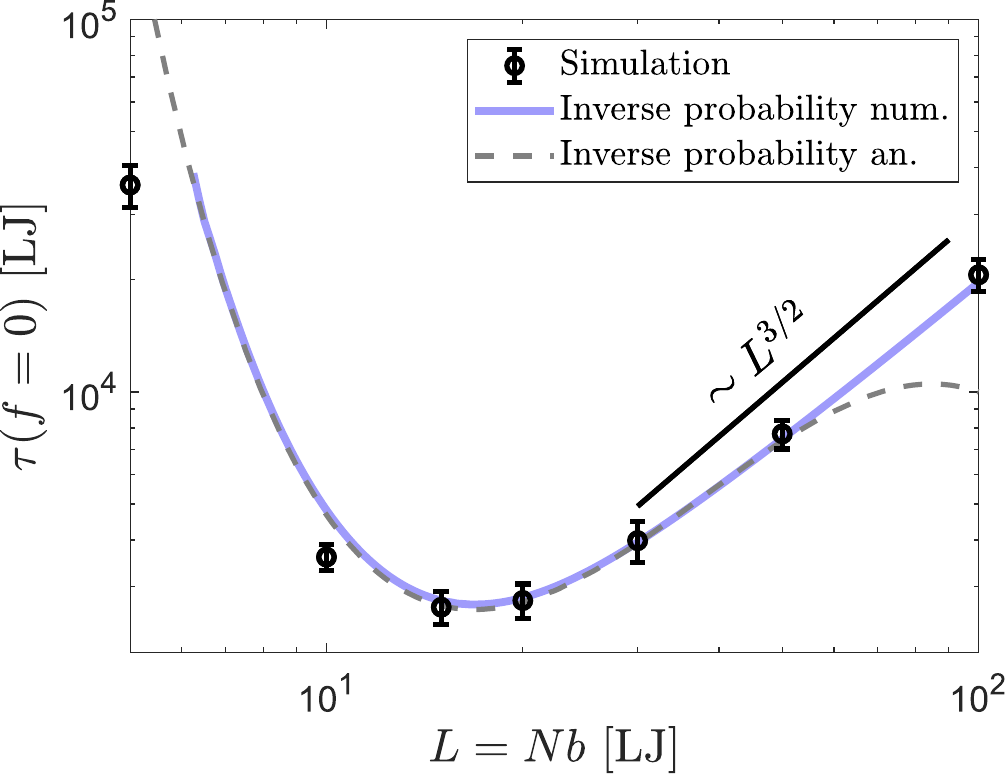}
\caption{Looping time at zero force as measured in a simulation for a semiflexible chain (symbols), compared to $1/P_\text{loop}(f=0)$. The full purple line gives the numerical result obtained using the algorithm of Ref.~\onlinecite{sinha2017ring}, the dashed grey line is the analytical result of Ref.~\onlinecite{shimada1984ring} valid for short chains. Both the purple and grey lines are shifted arbitrarily as to match simulation data well, as they only provide a scaling. For long chains, the same scaling as for a freely jointed chain is expected, resulting in~\cite{jun2003diffusion, laeremans2024looping} $\sim L^{3/2}$. In dimensionless units (Lennard-Jones), the simulation parameters are taken to be $\beta = b = r_\mathrm{c} = \gamma = 1$, $K = 100$ and $\kappa = 5$.}
\label{fig:zero_force} 
\end{figure}

An extra asset of the inverse scaling with the looping probability, is that it can be used as well to explore the dependence of the looping time on other parameters like the length of the DNA molecule, which was noted before in Ref.~\onlinecite{jun2003diffusion}. In Fig.~\ref{fig:zero_force}, we illustrate this by comparing simulation data for $\tau(f = 0)$ with the inverse looping probability, ranging over different lengths $L$. We find again excellent agreement for long chains. The equilibrium looping probability, we found numerically using the algorithm of Ref.~\onlinecite{sinha2017ring} (full purple line). For a stiff semiflexible chain in the continuous limit, the probability distribution $P(\vec{r})$ is given by~\cite{sinha2017ring}
\begin{align}
    P(\vec{r}) &= \frac{1}{Z} \int_{0}^{L} \ \mathcal{D}\left[ \hat{t}(s) \right] e^{-\frac{\beta l_P}{2} \int_{0}^{L} \left( \frac{d\hat{t}}{ds} \right)^2 ds} \delta^3\left(\vec{r} - \int_{0}^{L} \hat{t} \ ds \right),  \label{eq:P_full}
\end{align}
with in the discrete case $L = Nb$ the length of the polymer and $\vec{\hat{t}}(s)$ the unit tangent vector in function of the curvilinear distance $s$. Using the numerical scheme proposed in Ref.~\onlinecite{sinha2017ring}, one can find the looping probability and hence predict the looping time~\cite{jun2003diffusion, laeremans2024looping} as $\tau(f = 0) \sim 1/P_\text{loop}(f=0)$. We used as well the analytical result of Ref.~\onlinecite{shimada1984ring} valid for short chains (gray dashed line), reading~\cite{shimada1984ring, jun2003diffusion, guerin2017analytical}
\begin{align}
    P_\text{loop}(f=0)\Big\rvert_{L \ \text{small}} &\sim \frac{1}{L^5} e^{-14.055l_P/L + 0.246L/l_P}, \label{eq:P_approx}
\end{align}
For very long chains, a semiflexible chain can be mapped to a freely jointed chain, resulting in a looping probability that scales as~\cite{jun2003diffusion, laeremans2024looping}
\begin{align}
    P_\text{loop}(f=0)\Big\rvert_{L \  \text{large}} &\sim L^{-3/2}, \label{eq:P_approx2}
\end{align}
In Fig.~\ref{fig:zero_force}, these predictions are compared to simulation data, by which it is found that for $L$ large ($L \geq 15$), $\tau(f=0) \sim 1/P_\text{loop}(f=0)$ is indeed satisfied. For smaller $L$, the capture radius $r_\mathrm{c}$ becomes too large as compared to the total length for the inverse scaling to be valid. 

\subsection*{Blumberg free energy minimization}
\label{app:blum_free}
To find the minimum $\langle x(f) \rangle$ of the free energy as defined in Eq.~\ref{eq:blumfree} of the main text --- important for both the the two-state model (Eq.~\ref{eq:Blumbergeq}) and the inverse scaling with the looping probability (Eq.~\ref{eq:Wout}) --- we used the analytical approximation of Ref.~\onlinecite{petrosyan2017improved}, which reads
\begin{align}
    \frac{\langle x(f) \rangle}{L} &= \frac{4}{3} - \frac{4}{3\sqrt{\beta fl_P} + 1} - \frac{10 e^{\sqrt[4]{\frac{900}{\beta fl_P}}}}{\sqrt{\beta fl_P} \left( e^{\sqrt[4]{\frac{900}{\beta fl_P}}} - 1 \right)^2} \nonumber \\&+ \frac{\left( \beta fl_P \right)^{1.62}}{3.55 + 3.8 \left( \beta fl_P \right)^{2.2}}. \label{eq:force-ext}
\end{align}
Hence, both the two-state model and our inverse scaling with the looping probability can be evaluated fully analytically.

\begin{figure}[t]
    \centering
    \includegraphics[width=0.8\linewidth]{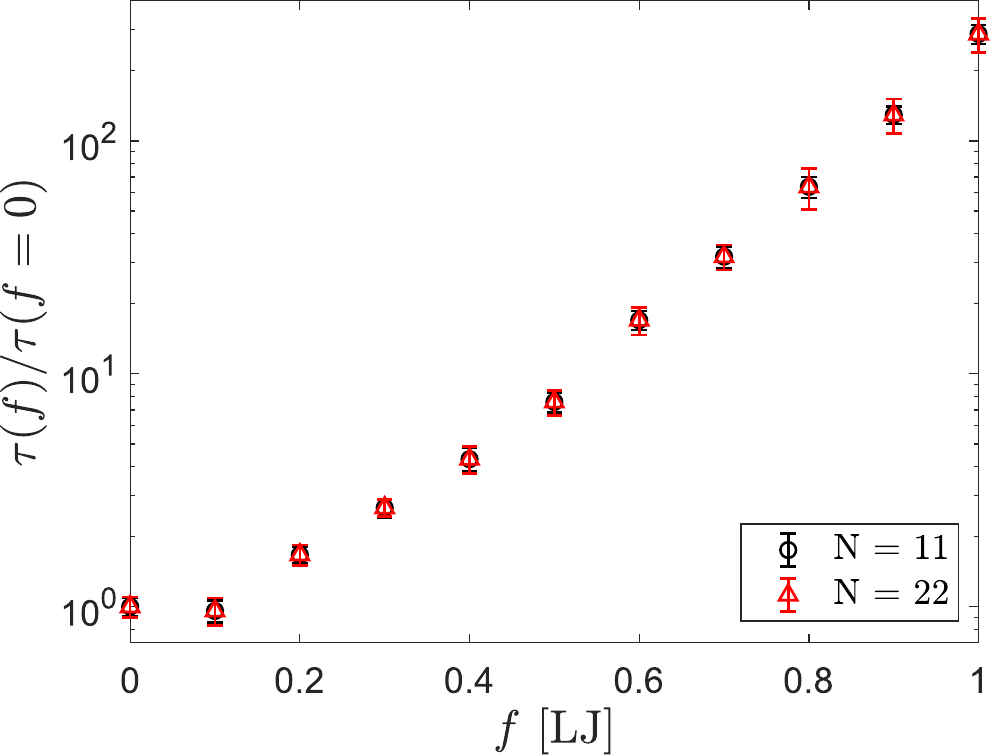}
\caption{Normalized looping time $\tau(f)/\tau(f=0)$ for $L = 11$ using two different coarse-grainings. The first (black triangle) has $N = 11$, $b = 1$ and $\kappa = 5$, while the second (red triangle) has $N = 22$, $b = 0.5$ and $\kappa = 10$. No noticeable difference can be seen, hence no finite size effects are present.}
\label{fig:fse} 
\end{figure}

\subsection*{Molecular dynamics simulations}
\label{app:comp_sim}
The simulations were performed using LAMMPS~\cite{LAMMPS}, where we took a bead-spring model with a quadratic spring energy and bending potential as explained in the main text. Before measuring the looping time, the starting configuration was taken to be a straight line with all springs in their rest length. All beads (having mass $m = 1$) received an initial velocity sampled from a Gaussian distribution with variance 1 and the chain was equilibrated for $10^5$ timesteps, with one timestep being $\Delta t = 0.05$. During equilibration, the dynamics was simulation using \textit{fix langevin}. Subsequently, we reduced the timestep to $\Delta t = 0.005$ and we used \textit{fix ffl} for the dynamics instead~\cite{hijazi2018fast}. Similar to Ref.~\onlinecite{laeremans2024looping}, we used REACTER~\cite{gissinger2020reacter} to keep track of when a loop was formed.

To define an error $\Delta$, we used the standard error of a sample mean, given by 
\begin{align}
    \Delta{\tau(f)} &= \frac{\sigma[\tau(f)]}{\sqrt{100}},
\end{align}
with $\sigma$ representing the sample standard deviation and $100$ being the number of independent simulations for each data point. In Fig.~\ref{fig:comp_sim}, we rescaled as $\tau(f) / \tau(f=0) \times C$, so we carry on the constant rescaling factor $C/\tau(f=0)$
\begin{align}
    \Delta \Big\{ \frac{C \times \tau(f)}{\tau(f=0)} \Big\} \approx \frac{C\Delta \tau(f)}{\tau(f=0)}.
\end{align}

\subsection*{Finite size effects}
In order to check whether the chain of length $N = 11$ suffers from finite size effects, we redid the simulations with a finer coarse-graining. Where originally we had $b = 1$ and $N = 11$, we now took $b = 0.5$ and $N = 22$. As the stiffness is given by $l_P = \beta \kappa b = 5$, we now have $\kappa = 10$. The result is shown in Fig.~\ref{fig:fse}, from which it is clear there is no noticeable difference.

\end{document}